\documentclass{JHEP3}
\usepackage{amsmath,amssymb}
\usepackage{multirow}
\usepackage{bm}
\usepackage{epsfig}
\usepackage{graphicx}
\usepackage{amsmath}
\usepackage{yfonts}
\usepackage{subfigure}

\makeatletter \@addtoreset{equation}{section}

\newcommand{\be}{\begin{equation}}
\newcommand{\ee}{\end{equation}}
\newcommand{\ba}{\begin{eqnarray}}
\newcommand{\ea}{\end{eqnarray}}

\newcommand{\bsubeq}{\begin{subequations}}
\newcommand{\esubeq}{\end{subequations}}

\def\lsim{\mathrel{\rlap{\lower3pt\hbox{\hskip1pt$\sim$}}
     \raise1pt\hbox{$<$}}} 
\def\gsim{\mathrel{\rlap{\lower3pt\hbox{\hskip1pt$\sim$}}
     \raise1pt\hbox{$>$}}}
\def\HLS1{HLS$_1$}

\DeclareFontFamily{U}{rsf}{}
\DeclareFontShape{U}{rsf}{m}{n}{
  <5> <6> rsfs5 <7> <8> <9> rsfs7 <10-> rsfs10}{}
\DeclareMathAlphabet\Scr{U}{rsf}{m}{n}

\allowdisplaybreaks

\title{Equations of state and compact stars in gauge/gravity duality}

\author{Kyung Kiu Kim
\\ Department of Physics, Kyung Hee University, Seoul 130-701, Korea
\\ Center for Quantum Spacetime, Sogang University, Seoul 121-742, Korea
\\ E-mail: \email{kimkyungkiu@gmail.com}}

\author{Youngman Kim
\\ Asia Pacific Center for Theoretical Physics and Department of Physics, Pohang University of Science and Technology, Pohang, Gyeongbuk 790-784, Korea
\\  Institute for Basic Science, Daejeon 305-811, Korea
\\ E-mail: \email{ykim@ibs.re.kr}}

\author{Ik Jae Shin
\\ Asia Pacific Center for Theoretical Physics, Pohang, Gyeongbuk 790-784, Korea
\\ Institute for Basic Science, Daejeon 305-811, Korea
\\ E-mail: \email{geniean@ibs.re.kr}}

\abstract{
We propose a new doorway to study the interplay between equations of state of dense matter and compact stars in gauge/gravity correspondence.
For this we construct a bulk geometry near the boundary of five-dimensional spacetime.
By solving a constraint equation derived from the bulk equation of motion together with the Tolman-Oppenheimer-Volkoff equation, we determine the equations of state for compact stars.  The input parameters in this study are the energy density and pressure at the center of the compact objects.
We also study how the equation of state depends on the parameters.
}


\begin{document}

\section{Introduction}
The equation of state (EOS) of dense matter finds its pivotal importance in various physical systems such as compact astrophysical objects and atomic nuclei.
Using a given EOS of dense (hadronic) matter, we can determine the mass and radius of a neutron star by solving the Tolman-Oppenheimer-Volkov (TOV) equations.
Or, using the observed mass and radius of a neutron star, one can constrain the EOS of dense matter \cite{Lattimer:2000nx, Ozel:2009da, Read:2008iy}.
Ultimately, one can explore the one-to-one correspondence between an EOS of dense matter and a mass-radius relation of a compact star.

Recently there have been several interesting attempts based on AdS/CFT \cite{Maldacena:1997re, Gubser:1998bc, Witten:1998qj} to understand compact stars \cite{deBoer:2009wk, Arsiwalla:2010bt, Parente:2010fs, Kim:2011da},  self-bound dense objects such as atomic nuclei \cite{Hashimoto:2008jq, Kim:2010an, Hashimoto:2011nm, Pahlavani:2011zzb}, and a two-dimensional plasma ball~\cite{Emparan:2009dj}.

In this work, we try to understand the compact stars in gauge/gravity duality.
Since one essential feature associated with the compact star is the gravitational force,
one has to construct field theory coupled to gravity.
However, the AdS/CFT is achieved by taking the decoupling limit where the gravity is completely decoupled from a boundary field theory system.
Therefore, one has to properly introduce the boundary gravity to any constructions for gravitationally bound compact stars.
 Such a construction can be realized through an effective action
\begin{eqnarray}
\label{effective action}
S_{eff} = W_{CFT}[g_{(0)}] + \frac{1}{16\pi G_4} \int d^4 x \sqrt{- g_{(0)}} ~R [g_{(0)} ] + \ldots~,
\end{eqnarray}
where $W_{CFT}[g_{(0)}]$ is the generating functional of the connected Green's functions for conformal field theory (CFT) with the boundary metric $g^{(0)}_{\mu\nu}$ and $G_4$ is the four-dimensional Newton constant. The dots denote  higher derivative curvature terms.
 A well-established way to obtain an effective action of a boundary field theory system  coupled to gravity
  in AdS/CFT is to introduce a finite
 UV-cutoff to the CFT system\footnote{There is an alternative way suggested recently, where normalizable gravity wave functions are introduced by an IR-cutoff \cite{Kiritsis:2006}. }.
 The cutoff defines an effective four-dimensional Newton constant along with other parameters from five-dimensions.
 This generalization of the AdS/CFT correspondence has been studied in \cite{Gubser99,Hawking00}, see \cite{Tanaka11} for a recent review.
 We will provide this kind of example in appendix~\ref{Exam_1}.

With the generalized AdS/CFT (or gauge/gravity) duality at hand,
 we try to build up a holographic model in which conformal field theory interacts with the boundary gravity
 and possibly uphold a stellar object.
For this, we take a simple ansatz for the bulk metric and the bulk matter field. Then, we solve the bulk equations of motion near the boundary
in  Fefferman-Graham coordinates.
 Our strategy to deal with the holographic perfect fluid stars is to put perfect fluid star metric as a boundary metric
 and  find the corresponding bulk geometry near the boundary of five-dimensional spacetime.
 In this procedure, we obtain a constraint from the bulk equation of motion.
 The constraint together with the TOV equations will determine the EOS of a dense matter.
As it is, this observation stands quite opposite to a standard method in which an EOS is an input to solve the TOV equations.
Mathematically, it is quite obvious that we could fix the EOS with the additional constraint from the bulk.
However, physically it is not clear how the constraint from the extra dimension could affect the EOS in four-dimensional spacetime.
With our results at hand, we study the interplay between equations of state of dense matter and perfect-fluid stars and classify sets of our model parameters. The values of our model parameters are specified at the center of a the perfect-fluid star. We also consider a case in which we
impose boundary conditions at the center and at the surface of the perfect-fluid star.

This paper is organized as follows.
In section \ref{compactStar}, we briefly review compact stars.
In section \ref{uniform}, we consider the simplest case in a holographic approach, the pure gravity system in the bulk, and we explain our assumptions and ansatz for the perfect-fluid star.
In section \ref{non-uniform}, we generalize the simplest case by adding a bulk scalar field.
We then show that this modification makes more realistic configuration possible.
In section \ref{summary}, we make a brief summary of our present study and provide some future extensions.
In appendix~\ref{Exam_1}, we review a construction for conformal field theory coupled to gravity studied in \cite{Hawking00}.

\section{Compact stars in a nutshell} \label{compactStar}
\begin{figure}
\begin{center}
\includegraphics[scale=0.25]{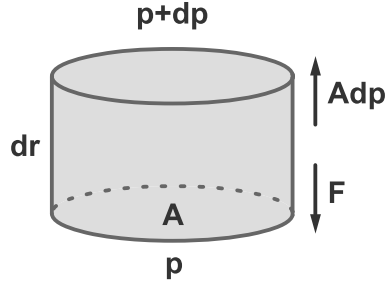}
\caption{Sketch for hydrostatic equilibrium. Here, $F$ denotes the gravitational force and $AdP$ is for the force from the thermal/degeneracy pressure.}
\label{fig_TOV}
\end{center}
\end{figure}

The formation of a (gravitationally bound) star is mainly governed by two forces:
 gravitation and thermal/degeneracy pressure generated by the matter inside the star.
In the case of compact stars, the degeneracy pressure is dominated by neutrons (neutron stars) and by electrons (white dwarfs).
Note that the energy of the white dwarfs is dominated by nuclei.
The force balancing between the gravitational collapse and the internal pressure leads to a basic structure equation for stars, see figure \ref{fig_TOV},
\ba
A\, dP= -\frac{G_4 m(r)\, dm}{r^2}\, ,
\ea
 where $AdP$ denotes a net force on the mass element due to the pressure difference $dP$.
 Since $dm=\rho(r)A dr$,  we obtain the equation of hydrostatic equilibrium, for instance see \cite{DMaoz},
\ba
\frac{dP(r)}{dr} =-\frac{G_4 m(r)\rho(r)}{r^2}\, .\label{TOV_NR}
\ea
$m(r)$ is the mass interior to a radius $r$, $m(r)=\int_0^r \rho(r^\prime) 4\pi {r^\prime}^2 dr^\prime$.
From this, we derive
\ba
\frac{d m(r)}{d r}=4\pi r^2\rho(r)\, .\label{TOV2}
\ea
Note that we do not distinguish the matter density from the energy density ($=$ matter density $\times~c^2$) since we take $c=1$, where
$c$ is the speed of light.

The treatment in the above is valid in the Newtonian limit. A gravitational effect, however, is well-described by general relativity.
So we move onto a solution of the Einstein equation that is essential to study a static star with spherical symmetry, see \cite{MTW} for more details.
For a static spherically symmetric stellar object, the natural and obvious metric ansatz  is
\begin{equation} \label{SchCoord}
	ds^2=-e^{2f(r)}dt^2+e^{2h(r)}dr^2+r^2 d\Omega_2^2\, ,
\end{equation}
where $d\Omega_2^2=d\theta^2+\sin^2\!\theta\,d\phi^2$.
With this ansatz we solve the four-dimensional Einstein equation
\begin{equation}
	 R_{\mu\nu}-\frac{1}{2}g_{\mu\nu}R=8\pi G_4\,T_{\mu\nu}\,.
\end{equation}
Here, we make an assumption that the matter inside the star is perfect fluid.
Then, the energy-momentum tensor can be written as
\begin{equation}
	T^{\mu\nu}=(\rho+P)u^\mu u^\nu+P\,g^{\mu\nu}\, ,
\end{equation}
where $u^\mu$ is the 4-velocity of the fluid with the normalization: $u^\mu u_\mu=-1$.
From the ($t,t$)-component of the Einstein equation, we obtain
\begin{equation}
	\frac{1}{r^2}\frac{d}{dr}\left[r\left(1-e^{-2h(r)}\right)\right]=8\pi G_4\,\rho(r)\,.
\end{equation}
With $2 G_4 m(r)=r(1-e^{-2h(r)})$, we can rewrite this equation as
\begin{equation} \label{interiorMass}
	\frac{dm(r)}{dr}=4\pi r^2\,\rho(r)\,,
\end{equation}
which is the same with (\ref{TOV2}).
Then, the metric (\ref{SchCoord}) changes into
\begin{equation} \label{interiorMet}
	ds^2=-e^{2f(r)}dt^2+\left(1-\frac{2G_4\,m(r)}{r}\right)^{-1}dr^2+r^2 d\Omega_2^2\,.
\end{equation}
On the other hand, from the ($r,r$)-component of the Einstein equation, we obtain
\begin{equation} \label{constraintF}
	f'(r)=\frac{G_4\left(m(r)+4\pi r^3 P(r)\right)}{r\left(r-2G_4\,m(r)\right)}\,.
\end{equation}
The relativistic extension of hydrostatic equilibrium is derived from the conservation law of energy-momentum tensor.
\begin{equation}
	\nabla_\mu T^{\mu\nu}=0~~~~~\Rightarrow~~~~~P^\prime(r)=-\left(P(r)+\rho(r)\right)f^\prime(r)
\end{equation}
Inserting (\ref{constraintF}) into above equation, we arrive at the TOV equation
\ba
\frac{dP(r)}{dr}=-\frac{G_4\left(m(r)+4\pi r^3 P(r)\right)\left(P(r)+\rho(r)\right)}{r\left(r-2G_4 m(r)\right)}\, ,\label{TOV_GR}
\ea
which is the general relativistic extension of (\ref{TOV_NR}).
To solve the equations, (\ref{TOV_NR}) (or (\ref{TOV_GR})) and (\ref{TOV2}), one generally needs to impose two initial conditions at $r=0$, $m(0)=0$ and
$\rho(0)=\rho_c$, and the EOS, $P=P(\rho)$. Usually one has to solve the equations numerically,
but there are several analytic solutions.
The simplest analytic solution is the uniform density star, $\rho(r)=\rho_c$.
In this case, the corresponding inner pressure is given by, see \cite{MTW} for instance,
\begin{equation}
	P(r)=\rho_c\,\frac{\left(1-2G_4 M r^2/R^3\right)^{1/2}-\left(1-2G_4 M/R\right)^{1/2}}{3\left(1-2G_4 M/R\right)^{1/2}-\left(1-2G_4 M r^2/R^3\right)^{1/2}}\, .
\end{equation}
Another interesting analytic solution is so called Tolman VII solution:
\ba
\rho(r)=\rho_c \left ( 1-(r/R)^2\right)\, ,
\ea
where $\rho_c$ is the central density and $R$ is the radius of the star.
The corresponding pressure is given as an analytic form. Since it is too complicated, we will not show
the pressure here. For a summary of the analytic solutions, we refer to \cite{Lattimer:2000nx}.

Outside of the star, the density $\rho(r)$ and pressure $P(r)$ vanish since there is no matter.
Then, from the equation (\ref{interiorMass}) we can see that $m(r)$ becomes a constant.
When we integrate the equation (\ref{constraintF}) with the condition $P(r)=0$ and $m(r)=m(R)=M$,
it requires a boundary condition $f(\infty)=0$ since the metric should be asymptotically flat.
Then, the final form of the exterior metric is
\begin{equation} \label{exteriorMet}
	ds^2=-\left(1-\frac{2G_4\,M}{r}\right)dt^2+\left(1-\frac{2G_4\,M}{r}\right)^{-1}dr^2+r^2 d\Omega_2^2\,.
\end{equation}
This is known as the Schwarzschild metric. From now on, we will set $G_4=1$.

\section{Uniform density stars} \label{uniform}

We start by considering the five-dimensional vacuum Einstein equation with a negative cosmological constant.
 \begin{equation} \label{bulk einstein eq}
	R_{mn}-\frac{1}{2}\mathcal{G}_{mn}R-\frac{6}{L^2}\mathcal{G}_{mn}=0
 \end{equation}
From the result in \cite{Skenderis:2000}, one may bring the general behavior of the bulk metric in the Feffermann-Graham coordinate as
 \begin{equation}
	ds^2=\mathcal{G}_{mn}dx^m dx^n=\frac{L^2}{z^2}\left(dz^2+g_{\mu\nu}(x,z)dx^\mu dx^\nu\right)\,,
 \end{equation}
where
 \begin{equation}
	g_{\mu\nu}(x,z)=g^{(0)}_{\mu\nu}(x)+z^2 g^{(2)}_{\mu\nu}(x)+z^4 g^{(4)}_{\mu\nu}(x)+\sum_{n=5}^{\infty}z^n g^{(n)}_{\mu\nu}(x)\,. \label{metric ansatz}
 \end{equation}
 We confirmed that the coefficients of $z$ and $z^3$ are identically zero.
In general, there can be a logarithmic term in this expansion. This logarithmic term contributes to a scheme dependent part of the boundary energy-momentum tensor, and we will neglect this term for simplicity. Mathematically, this choice restricts $g_{\mu\nu}(x,z)$ to be the polynomial function of $z$ near the boundary of the AdS space.
  Since we are going to describe a static spherically symmetric stellar configuration, we take the following ansatz for the metric.
 \begin{equation}
	g_{\mu\nu}(x,z)dx^\mu dx^\nu=g_{tt}(r,z)dt^2+g_{rr}(r,z)dr^2+g_{\theta\theta}(r,z)\left(d\theta^2+\sin^2\theta d\phi^2\right)\,,
 \end{equation}
where $r$ is the radial coordinate of the boundary space.

\begin{figure}[ht]
 \center
	\includegraphics[width=7.2cm]{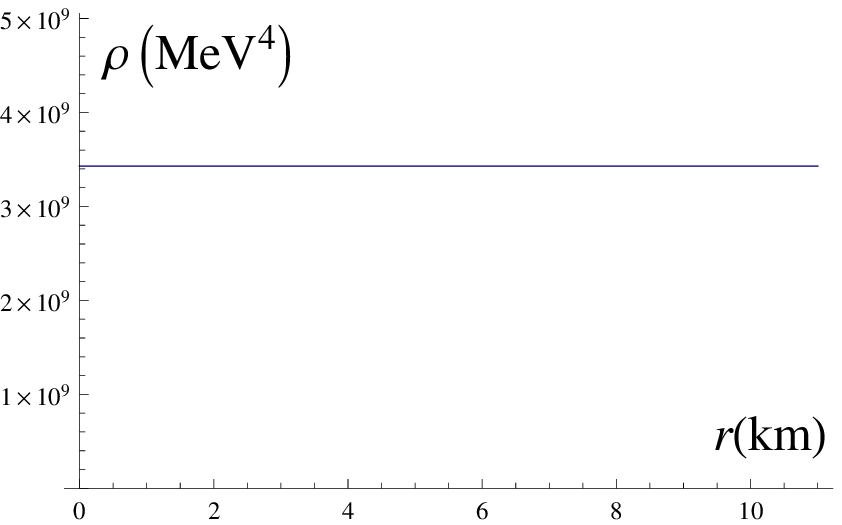}
	\includegraphics[width=7.2cm]{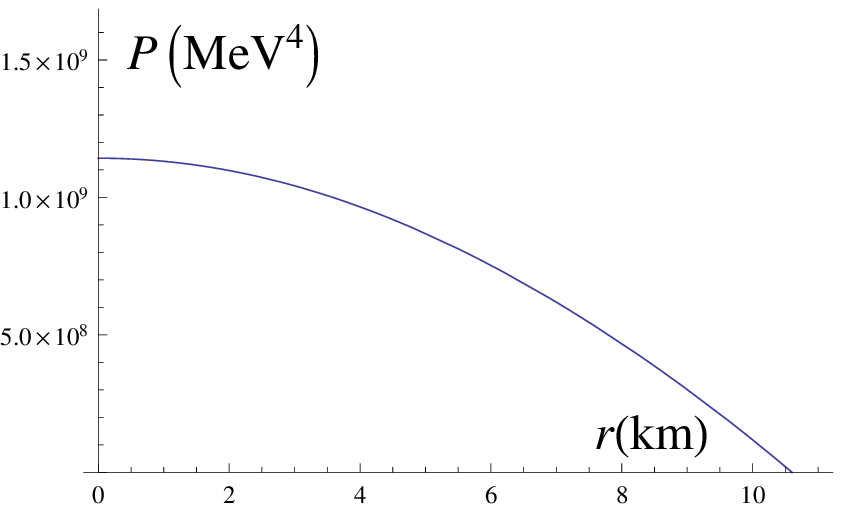}
	\caption{\small Energy density and pressure of a uniform density star}
	\label{f1}
 \end{figure}

Now, we are ready to solve the bulk Einstein equation (\ref{bulk einstein eq}) near the boundary.
Putting the metric ansatz into the Einstein equation,  we obtain differential equations of $g_{\mu\nu}^{(n)}$ with respect to $r$.
In the leading order of the expansion near the boundary, ${\cal O}(z^0)$, all the non-trivial components of $g_{\mu\nu}^{(2)}$ can be determined in terms of $g_{\mu\nu}^{(0)}$
 \begin{equation}
g_{\mu\nu}^{(2)}=-\frac{1}{2}\left(R_{\mu\nu}^{(0)}
-\frac{1}{6}R^{(0)}g_{\mu\nu}^{(0)}\right)\label{g2 from g0}\, .
 \end{equation}
 This is the same result with the one  in \cite{Skenderis:2000,Henningson:1998gx}.
 The equations obtained in the next-to-leading order, ${\cal O}(z^1)$, are trivially satisfied with the relation in (\ref{g2 from g0}).
Since we are interested in
a starlike object of the boundary system, we need to deform the boundary metric.
Without loss of generality, we consider the static spherically symmetric metric discussed in the previous section,
 \begin{equation}
	g^{(0)}_{tt}=-e^{2 f(r)}~,~~g^{(0)}_{rr}=\left(1-\frac{2m(r)}{r}\right)^{-1}~~~{\textrm {and}}~~~~g^{(0)}_{\theta\theta}=r^2\,.
 \end{equation}
Using this ansatz, the expression in (\ref{g2 from g0}) is given as follows.
\begin{eqnarray}
g_{tt}^{(2)}(r)&=&-\frac{e^{2f(r)}}{3r^2}\left(m'(r)-\left(3m(r)+r\left(-2+m'(r)\right)\right)f'(r)\right. \nonumber \\
&&~~~~~~~~~~~~~~~~~~~~~~~~~~~~~\left.+r\left(r-2 m(r)\right)f'(r)^2+r(r-2m(r))f''(r)\right) \\
g_{rr}^{(2)}(r)&=&\frac{1}{3r^2\left(r-2m(r)\right)}\left(3m(r)-2r m'(r)-r\left(r-3m(r)+r m'(r)\right)f'(r)\right. \nonumber \\
&&~~~~~~~~~~~~~~~~~~~~~~~~~~~~~~~~~~~~~~~~~~\left.+r^2\left(r-2m(r)\right)\left(f'(r)^2+f''(r)\right)\right) \\
g_{\theta\theta}^{(2)}(r)&=&-\frac{1}{6r}\left(3m(r)+r m'(r)-r\left(r-3m(r)+r m'(r)\right)f'(r)\right. \nonumber \\
&&~~~~~~~~~~~~~~~~~~~~~~~~~~~~~~~~~~~~~~~~~~\left.+r^2\left(r-2m(r)\right)\left(f'(r)^2+f''(r)\right)\right)
\end{eqnarray}
Considering the equations in the next-to-next-to-leading order, ${\cal O}(z^2)$, we obtain a constraint for the boundary metric.
The constraint has a little bit complicated form:
 \begin{eqnarray} \label{2ndGrr}
	&&0=-3m(r)^2+r\left(-4r+6 m(r)+rm'(r)\right)m'(r)+2r^2 \left(r-2 m(r)\right)m''(r)-2r\left(2r^2\right. \nonumber \\
	&&~~\left.+\left(-5r+3m(r)\right)m(r)+\left(-3r+4 m(r)+rm'(r)\right)m'(r)+2r^2\left(r-2m(r)\right)m''(r)\right)f'(r) \nonumber \\
	 &&~~+r^2\left(9r^2+\left(-32r+29m(r)\right)m(r)+r\left(-4r+6m(r)+rm'(r)\right)m'(r)\right. \nonumber \\
	 &&~~\left.+2r^2\left(r-2m(r)\right)m''(r)\right)f'(r)^2+2r^3\left(r-2 m(r)\right)\left(-3r+5m(r)+rm'(r)\right)f'(r)^3 \nonumber \\
	&&~~+r^4\left(r-2 m(r)\right)^2f'(r)^4-\left(4r^2\left(r-2 m(r)\right)\left(-r+m(r)+rm'(r)\right)\right. \nonumber \\
	&&~~\left.-4r^3\left(r-2 m(r)\right)\left(-m(r)+rm'(r)\right)f'(r)+2r^4\left(r-2 m(r)\right)^2f'(r)^2\right)f''(r) \nonumber \\
	&&~~+r^4\left(r-2 m(r)\right)^2f''(r)^2-2r^3\left(r-2 m(r)\right)^2\left(-1+rf'(r)\right)f'''(r) \,.
 \end{eqnarray}
In general, it is not easy to solve this equation since it is hard to impose suitable boundary conditions.
To make this equation in a more tractable form, we constrain our system further to consider {\em perfect fluid} stars.
In order to describe them, we put the following restriction on our metric as briefly summarized in the previous section, see \cite{MTW} for details.
 \begin{eqnarray}
	&&f'(r)=\frac{m(r)+4\pi r^3 P(r)}{r \left(r-2m(r)\right)}\,, \label{f'} \\
	&&m(r)=4\pi\int_0^r\rho (r'){r'}^2 dr'\,, \label{mass function} \\
	&&P'(r)=-(P(r)+\rho(r))\frac{m(r)+4\pi r^3 P(r)}{r\left(r-2m(r)\right)}\,, \label{TOV equation}
 \end{eqnarray}
where $\rho(r)$ and $P(r)$ are the energy density and pressure of a perfect fluid star.
Exploiting the above equations, we rewrite (\ref{2ndGrr}) in a much simpler form
 \begin{equation} \label{constraint 1}
	\rho '(r)=\frac{(P(r)+\rho(r))\left(3m(r)-4\pi r^3\rho(r)\right)}{r\left(r-3 m(r)-4\pi r^3 P(r)\right)}\,.
 \end{equation}
Now one can numerically solve (\ref{mass function}), (\ref{TOV equation}) and (\ref{constraint 1}) with the following boundary conditions: the value of the pressure and the energy density at the center of the perfect fluid star.
Note that $m(0)=0$ by construction.
An example of such solutions is given in figure \ref{f1}.
The example is a well-known uniform density star of the four-dimensional Einstein gravity in flat spacetime.
This configuration does not show interesting structure. However, it is not a meaningless result in the following sense.
This simple analysis demonstrates that our five-dimensional bulk construction near the boundary does constrain the four-dimensional gravity system, making it possible to determine the EOS and to study properties of the perfect fluid star.

Before we move to a realistic case, we discuss a bit more on the uniform density solution.
Though it is difficult (or almost impossible) to solve the coupled non-linear differential equations analytically,
we can study the behavior of the uniform density solution analytically near $r=0$ thanks to the regularity condition at the center.
From the relation (\ref{mass function}), $\rho(r)$  can be replaced with $m(r)$ as $\rho(r)=m'(r)/(4\pi r^2)$.
Furthermore, considering the regularity of $\rho(r)$ at $r=0$, one can impose the boundary conditions, $m'(0)=m''(0)=0$.
Now, we will expand $P(r)$ and $\rho(r)$ through  Taylor series around $r=0$.
Here, we focus on (\ref{constraint 1}),
\begin{eqnarray}
	0 &=& r\left(r-3m(r)-4\pi r^3 P(r)\right)\rho'(r)-(P(r)+\rho(r))\left(3m(r)-4\pi r^3\rho(r)\right) \nonumber \\
	&=& r\left(r-3m(r)-4\pi r^3 P(r)\right)\frac{d}{dr}\left(\frac{m'(r)}{4\pi r^2}\right)-\left(P(r)+\frac{m'(r)}{4\pi r^2}\right)\left(3m(r)-r m'(r)\right) \nonumber \\
	&=& \frac{m^{(4)}(0)}{24\pi}\,r^2 + \frac{m^{(5)}(0)}{48\pi}\,r^3 - \frac{\left(40\pi P(0)+5m^{(3)}(0)\right)m^{(4)}(0)-2m^{(6)}(0)}{320\pi}\,r^4 + \dots ~~~~~\, .
\end{eqnarray}
To satisfy the equation, it is clear that $m^{(4)}(0)=m^{(5)}(0)=0$, also $m^{(6)}(0)=0$.
Inserting these values into the original expression and performing the similar process repeatedly, we can see the higher-order derivative values of $m(r)$ at $r=0$ vanish except $m^{(3)}(0)=8\pi\rho(0)$.
This almost excludes possibility that all derivative terms of $\rho(r)$ do not vanish at $r=0$, and thus we may take $\rho(r)=\rho(0)=\textrm{const.}$
Here the word "almost" means that we didn't check if all the n-th higher-order derivative of  $m(r)$  vanish at $r=0$.


\section{Non-uniform density stars} \label{non-uniform}
Since the previous result is too simple to describe realistic compact stars, we try to improve the situation by introducing a bulk matter field.
 It is natural to expect that this matter field will lead non-uniform energy density since it will contribute to a $r$-dependent boundary action.
It is obvious from (\ref{TOV equation}) that the pressure will change (decrease) with the radius $r$.
For simplicity, we consider a bulk scalar\footnote{In a usual holographic QCD, this bulk scalar field
is dual to a quark bilinear operator.} and choose its bulk mass as $m_\phi^2=-3/L^2$.

\begin{figure}[ht]
 \center
	\includegraphics[width=4.9cm]{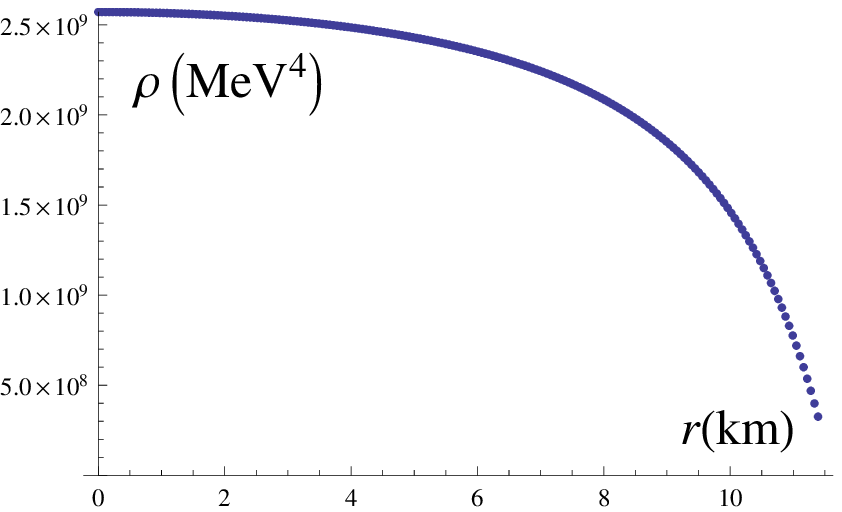}
	\includegraphics[width=4.9cm]{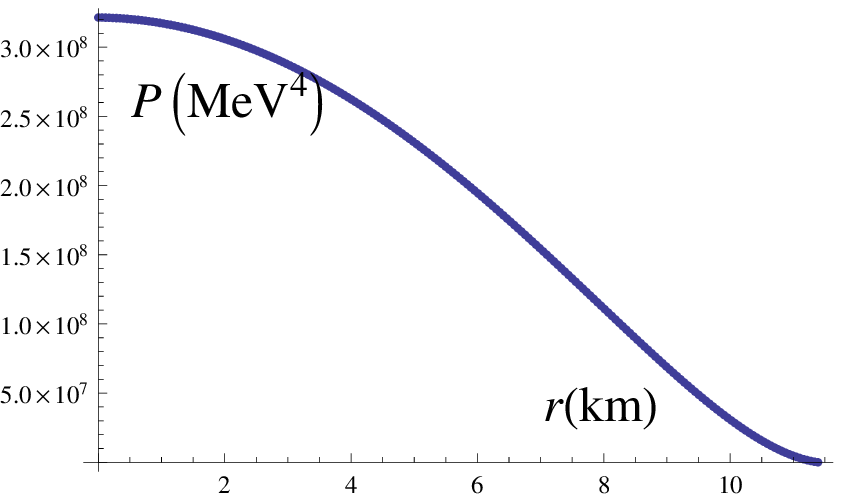}
	\includegraphics[width=4.9cm]{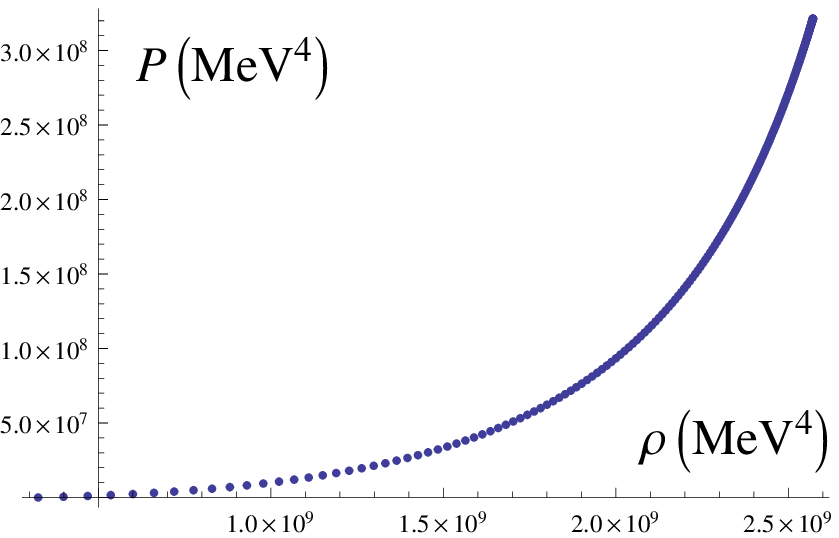}
	\caption{\small
		Energy density, pressure and equation of state for $\rho_c= 2.5713\times 10^9\,{\rm MeV}^4$ and $P_c=3.2141\times10^8\,{\rm MeV}^4$ with surface energy density $3.26\times10^8\,{\rm MeV}^4$.
		This configuration provides a star of the radius 11.45\,km and the mass 1.26\,$M_\odot$.}
	\label{first}
 \end{figure}

\begin{figure}[ht]
 \center
	\includegraphics[width=4.9cm]{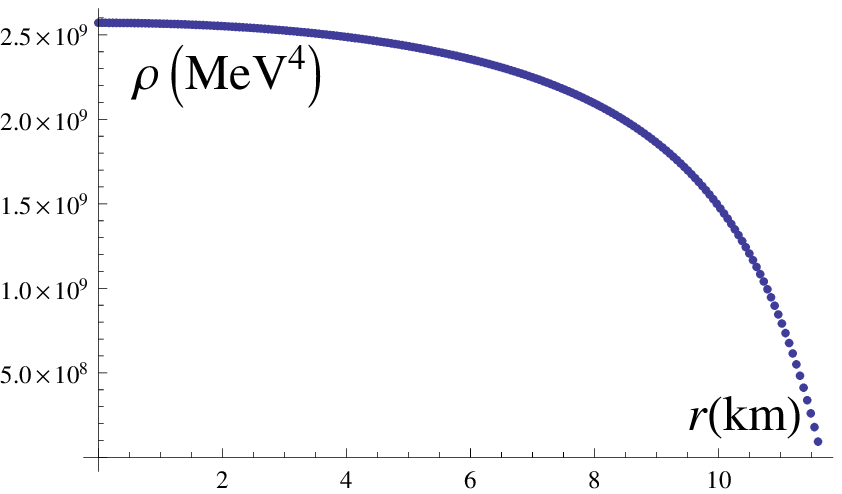}
	\includegraphics[width=4.9cm]{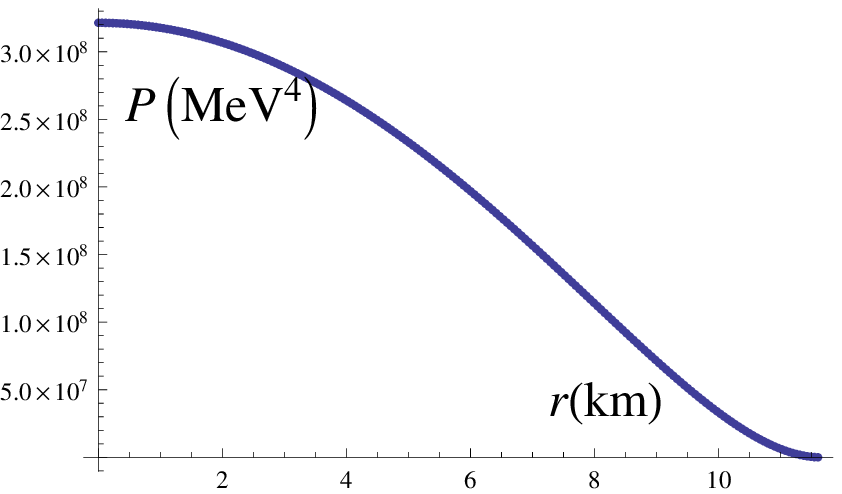}
	\includegraphics[width=4.9cm]{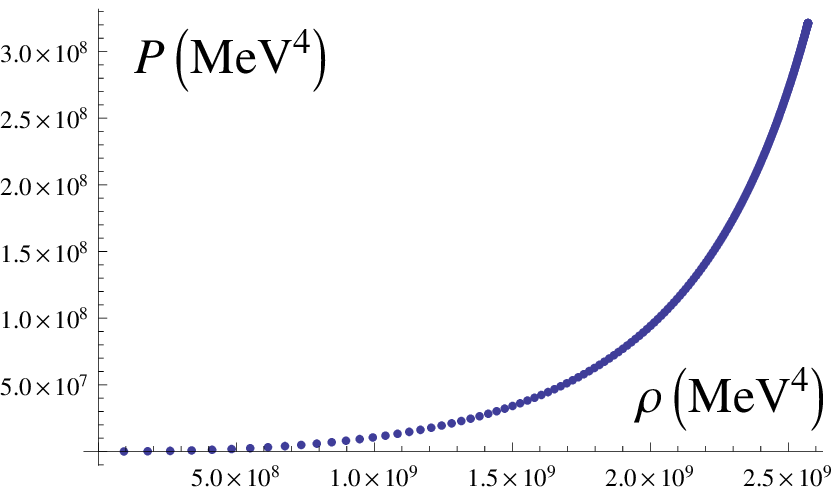}
	\caption{\small
		Energy density, pressure and equations of state for $\rho_c=2.5713\times10^9\,{\rm MeV}^4$ and $P_c=3.2141\times10^8\,{\rm MeV}^4$.
		This configuration gives the radius 11.66\,km and the mass 1.26\,$M_\odot$.}
	\label{Second}
 \end{figure}

The Einstein equation we have to solve is now
 \begin{eqnarray}
	&&R_{mn}-\frac{1}{2}\mathcal{G}_{mn}R-\frac{6}{L^2}\mathcal{G}_{mn}=T_{mn}\,, \label{EEq} \\
	 &&T_{mn}=\frac{1}{2}\partial_m\phi\,\partial_n\phi-\frac{1}{4}\mathcal{G}_{mn}\left(\left(\partial \phi\right)^2+m_\phi^2 \phi^2\right)\,.
 \end{eqnarray}
 Here we put the five-dimensional gravitation constant $G_5$ into the scalar field $\phi$.
In addition, we should solve the following equation of motion for the bulk scalar field
 \begin{equation}\label{scal EQ}
	(\nabla^2-m_\phi^2)\phi=0\,.
 \end{equation}
Now, we take an ansatz for the bulk scalar field.
 \begin{equation} \label{sEQ}
	\phi(r,z)=z\,\phi_1(r)+\sum_{n=3}^\infty z^n \phi_n(r)
 \end{equation}
 Similar to the metric expansion, the logarithmic  term is neglected. 
The form of the metric is the same with (\ref{metric ansatz}) near the boundary.
Using these ansatz, one can solve the equations, (\ref{EEq}) and (\ref{scal EQ}), perturbatively for small $z$.
For the matter part, we obtain (\ref{eq for mq}) for $\phi_{1}$ and (\ref{holo eos}) instead of (\ref{2ndGrr}).
 \begin{eqnarray}
	 &&0=r^2\phi_1(r)^3+\left(2m'(r)+\left(-2r+3m(r)+rm'(r)-r\left(r-2m(r)\right)f'(r)\right)f'(r)\right. \nonumber \\
	&&~~\left.-r\left(r-2 m(r)\right)f''(r)\right)\phi_1(r)+3\left(-2r+3m(r)+rm'(r)-r\left(r-2m(r)\right)f'(r)\right)\phi_1'(r) \nonumber \\
	&&~~-3r\left(r-2 m(r)\right)\phi_1''(r) \label{eq for mq} \\
	&& \nonumber \\
	&&0=\frac{1}{2}r^6\phi_1(r)^4+2r^3 \left(m(r)-r\left(r-2 m(r)\right)f'(r)\right)\phi_1(r)^2 \nonumber \\
	&&~~-2r^4\left(r-2 m(r)\right)\left(2+rf'(r)\right)\phi_1(r)\phi_1'(r)-3r^5\left(r-2 m(r)\right)\phi_1'(r)^2 \nonumber \\
	&&~~-3m(r)^2+r\left(-4r+6 m(r)+rm'(r)\right)m'(r)+2r^2 \left(r-2 m(r)\right)m''(r)-2r\left(2r^2\right. \nonumber \\
	&&~~\left.+\left(-5r+3m(r)\right)m(r)+\left(-3r+4 m(r)+rm'(r)\right)m'(r)+2r^2\left(r-2m(r)\right)m''(r)\right)f'(r) \nonumber \\
	 &&~~+r^2\left(9r^2+\left(-32r+29m(r)\right)m(r)+r\left(-4r+6m(r)+rm'(r)\right)m'(r)\right. \nonumber \\
	 &&~~\left.+2r^2\left(r-2m(r)\right)m''(r)\right)f'(r)^2+2r^3\left(r-2 m(r)\right)\left(-3r+5m(r)+rm'(r)\right)f'(r)^3 \nonumber \\
	&&~~+r^4\left(r-2 m(r)\right)^2f'(r)^4-\left(4r^2\left(r-2 m(r)\right)\left(-r+m(r)+rm'(r)\right)\right. \nonumber \\
	&&~~\left.-4r^3\left(r-2 m(r)\right)\left(-m(r)+rm'(r)\right)f'(r)+2r^4\left(r-2 m(r)\right)^2f'(r)^2\right)f''(r) \nonumber \\
	&&~~+r^4\left(r-2 m(r)\right)^2f''(r)^2-2r^3\left(r-2 m(r)\right)^2\left(-1+rf'(r)\right)f'''(r) \label{holo eos}
 \end{eqnarray}
As we did in the previous section, we consider perfect fluid star configurations,
and so we adopt the relations in (\ref{f'}), (\ref{mass function}) and (\ref{TOV equation}).
Then, the two equations, (\ref{eq for mq}) and (\ref{holo eos}), become much simpler,
 \begin{eqnarray}
	 &&\phi_1''(r)=\frac{1}{3r\left(r-2m(r)\right)}(r^2\phi_1(r)^3-4\pi r^2\phi_1(r)(3P(r)-\rho(r)) \nonumber \\
	 &&~~~~~~~~~~~~~~~~~~~~~~~~~~~~~~~~~~~~~~~~~~~~-6\left(r-m(r)+2\pi r^3\left(P(r)-\rho(r)\right)\right)\phi_1'(r))\,, \label{final mq} \\
	&&\rho'(r)=\frac{1}{32\pi r(-r+3m(r)+4\pi r^3P(r))}\left(-r^2 \left(-r\phi_1(r)^4+8\phi_1(r)\phi_1'(r)+6r\phi_1'(r)^2\right.\right. \nonumber \\
	&&~~~~~~~~~~~~~~~\left.\left.+16\pi rP(r)\phi_1(r)\left(\phi_1(r)+r\phi_1'(r)\right)\right)\right.+32 \pi\left(-3m(r)+4\pi r^3P(r)\right)\rho(r) \nonumber \\
	&&~~~~~~~~~~~~~~~\left.+12m(r)\left(-8\pi P(r)+r\phi_1'(r)\left(\phi_1(r)+r\phi_1'(r)\right)\right)+128\pi^2r^3\rho(r)^2\right)\,. \label{final rho}
 \end{eqnarray}
To solve the equations, (\ref{mass function}), (\ref{TOV equation}), (\ref{final mq}), and (\ref{final rho}),
 we need to impose three boundary conditions at $r=0$: $\rho(0)=\rho_c,~P(0)=P_c$ and $\phi_1(0)$,\footnote{Since (\ref{final mq}) is the second order differential equation, we need to have one more boundary condition, $\phi'_1(0)$. However, we can express it in terms of the other three conditions
 due to the regularity condition at the center.} together with regularity condition at the center of the object.
Here $\phi_1(0)$  has no significant physical meaning at this stage, but it may be associated with the energy density on the surface of our spherical objects, see below for more on this.
An example of the solutions is given in figure \ref{first}, where we choose $\rho_c=2.5713\times10^9\,{\rm MeV}^4$,~$P_c=3.2141\times10^8\,{\rm MeV}^4$
and $\phi_1(0)=0.2703$. In order to have some idea about corresponding baryon number density at the center of the star,
we resort to the pressure-baryon density plot in \cite{Ozel:2010fw} and find that our chosen $P_c$ corresponds to  roughly
$3.5$ times the normal nuclear matter density. In  \cite{Ozel:2010fw}, based on the recently determined mass-radius relation of neutron stars,
 the pressure of cold dense matter is {\it measured}.
In this way, we can obtain various EOSs once we specify the three values at the center of the compact stellar object.
However, among the three, the value of $\phi_1(0)$ is entirely arbitrary.
To remove this arbitrariness and to reduce the number of input parameters, we assume
that the energy density at the surface  $\rho(R)$ is zero. In this case, we can read off the value of $\phi_1(0)$ from our solutions.
In general, the surface density is much smaller than the central density, and therefore our assumption is a plausible  leading order approximation.
Then, our parameter space becomes two dimensional parameterized by the central values of the energy density and the pressure $(\rho_c,P_c)$.
Though the physical meaning of our inputs is now clear, we still have an ambiguity since we have no rigorous way to relate $\rho_c$ to $P_c$.
For illustrative purposes, we choose $P_c=\alpha \rho_c$ with $\alpha=1/3,1/4,1/5,1/6,1/8$.
The resulting mass-radius relation is summarized in table $\ref{mr relation}$.
For every point in the parameter space, we can find the corresponding equation of state.
This means that a set of central  density and pressure defines a unique equation of state.
 \begin{table}[h]
 \begin{center}
 \begin{tabular}{|c|c|c|c|c|c|c|c|c|c|}
	\hline
	 \multirow{3}*{$P_c=\frac{1}{3}\rho_c$}&$\rho_c$&~~$3\tilde\rho$~~&~~$4\tilde\rho$~~&~~$5\tilde\rho$~~~&~~$6\tilde\rho$~~~&~~$7\tilde\rho$~~~&~~$8\tilde\rho$~~~&~~$9\tilde\rho$~~~&~$10\tilde\rho$~~ \\
	\cline{2-10}
	&Mass\,($M_\odot$) &2.43 &2.11 &1.89&1.72 &1.59 &1.49 &1.40 &1.33 \\
	\cline{2-10}
	&Radius\,(km) &13.33 &11.54 &10.32 &9.42 &8.73 &8.16 &7.70 &7.30 \\
	\hline
 \end{tabular}
 \begin{tabular}{|c|c|c|c|c|c|c|c|c|c|}
	\hline
	 \multirow{3}*{$P_c=\frac{1}{4}\rho_c$}&$\rho_c$&~~$3\tilde\rho$~~&~~$4\tilde\rho$~~&~~$5\tilde\rho$~~~&~~$6\tilde\rho$~~~&~~$7\tilde\rho$~~~&~~$8\tilde\rho$~~~&~~$9\tilde\rho$~~~&~$10\tilde\rho$~~ \\
	\cline{2-10}
	&Mass\,($M_\odot$) &2.09 &1.81 &1.62&1.48  &1.37 &1.28  &1.20  &1.14 \\
	\cline{2-10}
	&Radius\,(km) &13.04&11.29&10.10&9.22&8.53 &7.98 &7.53&7.14 \\
	\hline
 \end{tabular}
 \begin{tabular}{|c|c|c|c|c|c|c|c|c|c|}
	\hline
	 \multirow{3}*{$P_c=\frac{1}{5}\rho_c$}&$\rho_c$&~~$3\tilde\rho$~~&~~$4\tilde\rho$~~&~~$5\tilde\rho$~~~&~~$6\tilde\rho$~~~&~~$7\tilde\rho$~~~&~~$8\tilde\rho$~~~&~~$9\tilde\rho$~~~&~$10\tilde\rho$~~ \\
	\cline{2-10}
	&Mass\,($M_\odot$) &1.81 &1.57 &1.40 &1.28 &1.19 &1.11 &1.05 &0.99 \\
	\cline{2-10}
	&Radius\,(km) &12.69 &10.99 &9.83 &8.97 &8.31 &7.77 &7.32 &6.95 \\
	\hline
 \end{tabular}
 \begin{tabular}{|c|c|c|c|c|c|c|c|c|c|}
	\hline
	 \multirow{3}*{$P_c=\frac{1}{6}\rho_c$}&$\rho_c$&~~$3\tilde\rho$~~&~~$4\tilde\rho$~~&~~$5\tilde\rho$~~~&~~$6\tilde\rho$~~~&~~$7\tilde\rho$~~~&~~$8\tilde\rho$~~~&~~$9\tilde\rho$~~~&~$10\tilde\rho$~~ \\
	\cline{2-10}
	&Mass\,($M_\odot$) &1.59 &1.38 &1.23 &1.12 &1.04 &0.97 &0.92 &0.87  \\
	\cline{2-10}
	&Radius\,(km) &12.33 &10.68 &9.55 &8.72 &8.07 &7.55 &7.12 &6.75 \\
	\hline
 \end{tabular}
 \begin{tabular}{|c|c|c|c|c|c|c|c|c|c|}
	\hline
	 \multirow{3}*{$P_c=\frac{1}{8}\rho_c$}&$\rho_c$&~~$3\tilde\rho$~~&~~$4\tilde\rho$~~&~~$5\tilde\rho$~~~&~~$6\tilde\rho$~~~&~~$7\tilde\rho$~~~&~~$8\tilde\rho$~~~&~~$9\tilde\rho$~~~&~$10\tilde\rho$~~ \\
	\cline{2-10}
	&Mass\,($M_\odot$) &1.26 &1.09 &0.97 &0.89 &0.82 &0.77 &0.73 &0.69 \\
	\cline{2-10}
	&Radius\,(km) &11.66 &10.10 &9.03 &8.25 &7.64 &7.14 &6.73 &6.39 \\
	\hline
 \end{tabular}
 \caption{\small
	Sample mass-radius relations.
	Here $\tilde{\rho}=8.5704\times10^8\,{\rm MeV}^4$.}
 \label{mr relation}
 \end{center}
 \end{table}
For instance, we demonstrate the energy density, the pressure and the equation of state for $\rho_c=2.5713\times10^9\,{\rm MeV}^4$ and $P_c=3.2141\times10^8\,{\rm MeV}^4$ in figure \ref{Second}.
In this case, we obtain $\phi_1(0)=0.2704$ from our solutions.

Now we try to understand the physical roll of $\phi_1(0)$. If we change the value of $\phi_1(0)$ above, we will obtain a star configuration with different surface energy densities. Thus, with {\em given} values of the central density and pressure as the boundary condition,
 $\phi_1(0)$ is closely tied to the surface energy density. This does not necessarily imply that $\phi_1(0)$ is {\em physically} tied to the surface energy density. For instance, if we use the surface energy density and the central pressure as the boundary condition, then
 $\phi_1(0)$ will be related to the central density.
 Therefore, within our framework, it is hard to put some physical meaning on $\phi_1(0)$.
 In terms of the holographic QCD, one may say that  $\phi_1(0)$ is nothing but the quark mass at the center of the star by the AdS/CFT dictionary and
 $\phi_1(r)$ is an effective density-dependent quark mass since the density
 varies with the radius of the compact star.

Before closing this section, we give a remark on the EOS obtained in this study.
In a usual approach to calculate the mass-radius relation,
one adopts an EOS from various studies in flat spacetime with no general relativistic effects and then solve the TOV equation.
Though it is obvious, we remark here that this standard approach,
TOV equation with flat EOS, is perfectly consistent with general relativity.
In our case, we solved the TOV equation together with the constraint from the bulk to obtain the EOS.
In this sense we can argue that our EOS includes the effect of general relativity (or gravity).

 \section{Summary}\label{summary}
In this work, we investigated stellar configurations in gauge/gravity correspondence to study perfect fluid stars.
 For this, we have considered holographic dual of a boundary field theory system coupled to gravity.
 To retain the gravity degrees of freedom, we introduced a finite UV-cutoff that defines the four-dimensional Newton constant.
With the assumption that the energy density at the surface is zero, we determined the EOS and mass-radius relation of a compact star,
 once the central values of the energy density and pressure are specified.
An ambiguity in this case is that the central values of the energy density and pressure are not correlated each other.
In this study, the EOS is not an input to solve the TOV equation, but the output of our analysis together with the corresponding mass and radius
of a star.

So far, we analyzed the bulk metric only near the boundary.
To explore the full bulk metric, we have to solve partial differential equations with appropriate boundary conditions.
Our study could be very important information for the boundary data. In future, we will seek a complete solution for the bulk metric that covers whole bulk space. Another possible extension is to generalize our ansatz. Since the generalization can affect directly energy-momentum tensor, stars can have interesting structure even in the pure bulk gravity case. A study with a more general ansatz for the bulk metric is relegated to a forthcoming publication.


\section*{Acknowledgement}
Y.K. thanks Chang-Hwan Lee for helpful discussion on neutron stars.
K.K. is grateful to Piljin Yi for useful discussion.
Y.K. and I.J.Shin acknowledge the Max Planck Society(MPG), the Korea Ministry of Education, Science and
Technology(MEST), Gyeongsangbuk-Do and Pohang City for the support of the Independent Junior
Research Group at APCTP. This work was supported by the World Class University grant R32-2009-000-10130-0 (K.K.) and the National Research Foundation of Korea(NRF) grant funded by the Korea government(MEST) with grant No. 2010-0023121  (K.K.)  and  2011-0023230 (K.K.) and also through the Center for Quantum Spacetime(CQUeST) of Sogang University with grant number 2005-0049409 (K.K.).

\appendix{}

\section{AdS/CFT with gravity \label{Exam_1}}

In this section, we provide an explicit example of a generalization of the AdS/CFT correspondence with gravity.
In \cite{Gubser99,Hawking00}, it is shown that the well-known RS-II model \cite{Randall:1999vf} is dual to a system which is similar to (\ref{effective action}).
Here we present a compact review on the derivation in \cite{Hawking00}.

The AdS/CFT correspondence is defined by equivalence between the partition function of five-dimensional gravity theory and the partition function of four-dimensional conformal field theory as follows.
\begin{eqnarray}\label{gAdSCFT}
Z[\gamma] = \int d[\delta\mathcal{G}] e^{i S_{gravity} [\delta\mathcal{G};\gamma]} = \int d[\varphi] e^{i S_{CFT}[\varphi,\gamma]}\equiv e^{i W_{CFT}[\gamma]}~~,
\end{eqnarray}
where $\gamma$ is the induced metric at the boundary of AdS space and $W_{CFT}[\gamma]$ denotes the generating functional for connected Green's functions. $\delta\mathcal{G}$ and $\varphi$ are path-integral measures. The action for the five-dimensional gravity theory is given by
\begin{eqnarray}\label{S gravity}
S_{gravity}= \frac{1}{16\pi G_5}\int d^5x \sqrt{-\mathcal{G}}(R + 2 \Lambda) + \frac{1}{8 \pi G_5} \int d^4 x \sqrt{-\gamma}~K + S_1 + S_2 + S_3~,
\end{eqnarray}
where the five-dimensional metric can be decomposed into
\begin{eqnarray}
ds^2=\mathcal{G}_{mn}dx^m dx^n= \mathcal{G}_{\mu\nu}(x,z)dx^\mu dx^\nu + \frac{L^2 dz^2}{z^2}~~,
\end{eqnarray}
and the range of coordinate $z$ is $z_0 < z <\infty$ with $z_0\ll L$.
The UV-cutoff $\Lambda$ is related to $z_0~(\sim \frac{1}{\Lambda})$.
The first term of (\ref{S gravity}) is the the well-known Einstein-Hilbert action with the cosmological constant,
 and the second term is the Gibbons-Hawking term.
The other terms are introduced by holographic renormalization as follows.\footnote{Where $g^{(0)}_{\mu\nu}$ denotes  z-independent part of $(z^2/L^2) \mathcal{G}_{\mu\nu}$}
\begin{eqnarray}
&&S_1 = -\frac{3}{8\pi G_5 L} \int d^4 x \sqrt{-\gamma}\, ,\\
&&S_2 =-\frac{L}{32 \pi G_5} \int d^4 x \sqrt{-\gamma} R^{(\gamma)}\, ,\\
&&S_3 = -\frac{L^3 \log (z_0/L)}{64 \pi G_5} \int d^4 x \sqrt{-g_{(0)}}\left[ R_{\mu\nu}(g_{(0)}) R^{ \mu\nu}(g_{(0)}) -\frac{1}{3}{R(g_{(0)})}^2 \right]
\end{eqnarray}
These are counter terms needed to cancel divergence in the Einstein-Hilbert action and the Gibbons-Hawking term.
On the other hand,  in view of four-dimensional boundary theory, these terms can be interpreted as follows.
$S_1$ is just four-dimensional cosmological constant, and $S_2$ is four-dimensional Einstein-Hilbert action.
$S_3$ could be interpreted as higher derivative corrections with coupling that depends on the cutoff scale.
Actually, it is proportional to the conformal anomaly.

Now, let us turn to the RS-II model, which is defined as,
\begin{eqnarray}
S_{RS-II} = S_{EH} + S_{GH} + 2 S_1 + S_m\, ,
\end{eqnarray}
where $S_{EH}$ and $S_{GH}$ are the Einstein-Hilbert action with cosmological constant and the Gibbons-Hawking term as in the previous case.
In addition, a brane exists at $z=z_0$ and the $2 S_1$ gives its tension, $\frac{3}{4\pi G_5 L}$.
The model has confined matter on the brane at $z=z_0$, which is described by the action $S_m$.
Let's consider path-integral of this action. One can write down the partition function as follows.
\begin{eqnarray}
Z_{RS-II}[\gamma] &=& \int d[\delta\mathcal{G}] d[\delta \varphi] e^{i S_{RS-II}[\mathcal{G}_0 + \delta\mathcal{G},\delta \varphi]}\\\nonumber&=&e^{2i S_1[ \gamma]}  \int d[\delta\mathcal{G}] d[\delta \varphi] e^{ iS_{EH}[\mathcal{G}_0 + \delta\mathcal{G}]  + i S_{GH}[\mathcal{G}_0 + \delta\mathcal{G}] + i S_m[\delta \varphi ; \gamma]}
\\\nonumber&=&e^{2i S_1[ \gamma]}  \left(\int_{one-bulk} d[\delta\mathcal{G}] e^{ iS_{EH}[\mathcal{G}_0 + \delta\mathcal{G}]  + i S_{GH}[\mathcal{G}_0 + \delta\mathcal{G}] }\right)^2  \left( \int d[\delta\varphi] e^{i S_m[\delta \varphi ;  \gamma]}\right)
\end{eqnarray}
Here to arrive at the final expression, we took account of the $Z_2$ symmetry, two bulk-regions, of the RS-II model.
Using the fact that integration of $S_{EH}$ and $S_{GH}$ already appeared in the (\ref{gAdSCFT})\footnote{$Z[\gamma] = e^{iS_1 + iS_2 + iS_3}\int d[\delta \mathcal{G}] e^{iS_{EH} + iS_{GH}}  = e^{i W_{CFT}[\gamma]}$ }, we can express the partition function as follows
\begin{eqnarray}
Z_{RS-II}[\gamma] = e^{2i\left(W_{CFT}- S_2 -S_3\right)[\gamma]} \int d[\delta\varphi] e^{i S_m [\delta\varphi; \gamma]}~.
\end{eqnarray}
As mentioned earlier,
$S_m$ is the action for the matter field confined on the brane in the RS-II model,
thus if we ignore the matter action, the extremum of the RS-II model is described by following effective action,
\begin{eqnarray}
S_{eff} = W_{CFT}[\gamma] - S_2[\gamma] - S_3[\gamma]\label{effective action 1}~.
\end{eqnarray}
Now, it is obvious from this equation that the RS-II model is dual to the conformal field theory coupled to gravity including higher curvature interaction. For more clear application to our paper, we use the result of \cite{Skenderis:2000}. Solving Einstein equation in five-dimensional AdS space, the above induced metric $\mathcal{G}_{\mu\nu}(x,z)$ can be decomposed as follows.
\begin{eqnarray}\label{metric expantion}
\mathcal{G}_{\mu\nu}= \frac{L^2}{z^2} \left( g^{(0)}_{\mu\nu}(x) +z^2 g^{(2)}_{\mu\nu}(x)  + z^4 \log z^2 h_{\mu\nu}(x)+\sum_{n=4}^{\infty}z^n g_{\mu\nu}^{(n)}(x)  \right)
\end{eqnarray}
If $g_{\mu\nu}^{(0)}$ and $g_{\mu\nu}^{(4)}$ are given, the other terms are determined. Here $g_{\mu\nu}^{(0)}$ is interpreted as the boundary metric at the leading order in $\frac{z_0}{L}$ and $g_{\mu\nu}^{(4)}$ is closely related to boundary energy-momentum tensor. Then we may write down above effective action in terms of the boundary metric $g_{\mu\nu}^{(0)}$ as follows.
\begin{eqnarray}
S_{eff}  &\cong& W_{CFT}[g_{(0)}] + \frac{L^3}{32 \pi G_5 z_0^2} \int d^4 x \sqrt{-g_{(0)}} R[g_{(0)}] \label{effective action 2}\\\nonumber&&+\frac{L^3 \log (z_0/L)}{64\pi G_5} \int d^4 x \sqrt{-g_{(0)}}\left(R_{\mu\nu}[g_{(0)}]R^{\mu\nu}[g_{(0)}]-\frac{1}{3} R[g_{(0)}]^2\right)+ \ldots~,
\end{eqnarray}
where ``$\ldots$" denotes higher curvature terms from $S_2[\gamma]$.

Let us consider the four-dimensional Einstein equation derived from this effective action. After variation, we can easily expect the following result
\begin{eqnarray}\label{4d einstein eq}
R_{\mu\nu}[g_{(0)}] - \frac{1}{2} g^{(0)}_{\mu\nu} R[g_{(0)}]=8\pi G_4 \left( T^{CFT}_{\mu\nu} + H_{\mu\nu} + \ldots~\right)~,
\end{eqnarray}
where $T^{CFT}_{\mu\nu}$ and $H_{\mu\nu}$ come from the variations of $W_{CFT}$ and $\log z_0$ term with respect to $g_{\mu\nu}^{(0)}$ and
 ``$\ldots$" means the contribution coming from the higher derivative correction terms. Thus we may regard the right hand side of (\ref{4d einstein eq}) as an effective energy momentum tensor $T_{\mu\nu}^{eff} = T_{\mu\nu}^{CFT} + H_{\mu\nu}+ \ldots$ , where $H_{\mu\nu}$ is proportional to $h_{\mu\nu}(x)$ in (\ref{metric expantion}).

So far, we have reviewed part of \cite{Hawking00} and translated their results in our convention.
From the result, we can read four-dimensional gravitational constant as $G_4=2G_5 z_0^2/L^3$ from (\ref{effective action 2}).
     Note that in the limit $z_0 \rightarrow 0$,  $G_4$ vanishes and gravity is decoupled from the field theory system.



\begin{thebibliography}{999}



\bibitem{Lattimer:2000nx}
  J.~M.~Lattimer and M.~Prakash,
  Astrophys.\ J.\  {\bf 550} (2001) 426  [astro-ph/0002232].

\bibitem{Ozel:2009da}
  F.~Ozel and D.~Psaltis,
  Phys.\ Rev.\ D {\bf 80} (2009) 103003  [arXiv:0905.1959 [astro-ph.HE]].

   \bibitem{Read:2008iy}
  J.~S.~Read, B.~D.~Lackey, B.~J.~Owen and J.~L.~Friedman,
  Phys.\ Rev.\ D {\bf 79} (2009) 124032  [arXiv:0812.2163 [astro-ph]].



 \bibitem{Maldacena:1997re}
	J.~M.~Maldacena,
	Adv.\ Theor.\ Math.\ Phys.\  {\bf 2 } (1998)  231-252.
	[hep-th/9711200].

 \bibitem{Gubser:1998bc}
	S.~S.~Gubser, I.~R.~Klebanov and A.~M.~Polyakov,
	Phys.\ Lett.\  B {\bf 428} (1998) 105
	[arXiv:hep-th/9802109].

 \bibitem{Witten:1998qj}
	E.~Witten,
	Adv.\ Theor.\ Math.\ Phys.\  {\bf 2} (1998) 253
	[arXiv:hep-th/9802150].

 \bibitem{deBoer:2009wk}
	J.~de Boer, K.~Papadodimas and E.~Verlinde,
	JHEP {\bf 1010} (2010) 020
	[arXiv:0907.2695 [hep-th]].
		
 \bibitem{Arsiwalla:2010bt}
	X.~Arsiwalla, J.~de Boer, K.~Papadodimas and E.~Verlinde,
	JHEP {\bf 1101} (2011) 144
	[arXiv:1010.5784 [hep-th]].

 \bibitem{Parente:2010fs}
	V.~Parente and R.~Roychowdhury,
	JHEP {\bf 1104} (2011) 111
	[arXiv:1011.5362 [hep-th]].

 \bibitem{Kim:2011da}
	Y.~Kim, C.~-H.~Lee, I.~J.~Shin and M.~-B.~Wan,
	JHEP {\bf 1110} (2011) 111
	[arXiv:1108.6139 [hep-ph]].

 \bibitem{Hashimoto:2008jq}
	K.~Hashimoto,
	Prog.\ Theor.\ Phys.\  {\bf 121} (2009) 241
	[arXiv:0809.3141 [hep-th]].
	
 \bibitem{Kim:2010an}
	K.~K.~Kim, Y.~Kim and Y.~Ko,
	JHEP {\bf 1010} (2010) 039
	[arXiv:1007.2470 [hep-ph]].

 \bibitem{Hashimoto:2011nm}
	K.~Hashimoto and T.~Morita,
	Phys.\ Rev.\ D {\bf 84} (2011) 046004
	[arXiv:1103.5688 [hep-th]].

 \bibitem{Pahlavani:2011zzb}
	M.~R.~Pahlavani, J.~Sadeghi and R.~Morad,
	J.\ Phys.\ G G {\bf 38} (2011) 055002.

\bibitem{Emparan:2009dj}
  R.~Emparan and G.~Milanesi,
  JHEP {\bf 0908} (2009) 012  [arXiv:0905.4590 [hep-th]].

 \bibitem{Kiritsis:2006}
	E.~Kiritsis and F.~Nitti,
	Nucl.\ Phys.\ B {\bf 772} (2007) 67
	[hep-th/0611344].
	
 \bibitem{Gubser99}
	S.~S.~Gubser,
	Phys.\ Rev.\ D {\bf 63} (2001) 084017
	[hep-th/9912001].

 \bibitem{Hawking00}
	S.~W.~Hawking, T.~Hertog and H.~S.~Reall,
	Phys.\ Rev.\ D {\bf 62} (2000) 043501
	[hep-th/0003052].

 \bibitem{Tanaka11}
	N.~Tanahashi and T.~Tanaka,
	Prog.\ Theor.\ Phys.\ Suppl.\  {\bf 189} (2011) 227
	[arXiv:1105.2997 [hep-th]].


\bibitem{DMaoz}
D. Maoz, {\em Astrophysics in a Nutshell},  Princeton University Press, 2007.

\bibitem{MTW}
C. W. Misner, K. S. Thorne, and J. A. Wheeler, {\em GRAVITATION}, W. H. FREEMAN AND COMPANY, New York.

 \bibitem{Skenderis:2000}
	S.~de Haro, S.~N.~Solodukhin and K.~Skenderis,
	Commun.\ Math.\ Phys.\  {\bf 217} (2001) 595
	[hep-th/0002230].

 \bibitem{Henningson:1998gx}
	M.~Henningson and K.~Skenderis,
	JHEP {\bf 9807} (1998) 023
	[hep-th/9806087].
	
 \bibitem{Ozel:2010fw}
	F.~Ozel, G.~Baym and T.~Guver,
	Phys.\ Rev.\ D {\bf 82} (2010) 101301
	[arXiv:1002.3153 [astro-ph.HE]].

 \bibitem{Randall:1999vf}
	L.~Randall and R.~Sundrum,
	Phys.\ Rev.\ Lett.\  {\bf 83} (1999) 4690
	[hep-th/9906064].
	
\end{thebibliography}
\end{document}